\documentstyle[epsf,sprocl]{article}
\def\Journal#1#2#3#4{{#1} {\bf #2}, #3 (#4)}\vspace*{-.1cm}
\def\fluss#1{$10^{#1}$ cm$^{-2}$s$^{-1}$}
\def\ea#1#2{#1 {et al.}, #2}

\def\apj{{\em Astrophys. Journal}}
\def\apjl{{\em Astrophys. Journal Lett.}}
\def\apjs{{\em Astrophys. Journal Suppl.}}
\def\app{{\em Astropart. Phys.}}
\def\apss{{\em Astrophys. \& Sp. Sc.}}
\def\araa{{\em Annu. Rev. Astron. Astrophys.}}

\def\hsv{{\em AIP Proc. 3rd Huntsville Symp. on GRB}, 1996 in press.}

\def\nat{{\em Nature}}
\def\NCA{{\em Nuovo Cimento}} 
\def\NIM{{\em Nucl. Instrum. Methods}}

\def\padova{{\em Proc. Towards a major atmospheric Cherenkov detector IV, Padova}}

\def\prs{{\em Phys. Rep.}} 

\def\rome{{\em Proc. 24th ICRC, Rome}}

\def\ssr{{\em Sp. Sc. Rev.}}
 
\def\et{{\em et al.,}} 
\def\be{\begin{equation}} 
\def\ee{\end{equation}}
\def\bea{\begin{eqnarray}} 
\def\eea{\end{eqnarray}}

\def\uka { \raisebox{-0.5ex} {\mbox{$\stackrel{<}{\scriptstyle \sim}$}}}

\def\deg {$^\circ$}
\begin{document}
\title{Search for TeV Counterparts of Gamma-Ray Bursts with the HEGRA Experiment
\vspace*{-0.55cm}
}
\author{Henric Krawczynski}
\address
{Universit\"at Hamburg, II. Inst. f\"ur Experimentalphysik, Luruper Chaussee 149,\\
        22761 Hamburg, Germany, email: krawcz@mail.desy.de
\vspace*{-0.14cm}
}
\author{Burkhart Funk}
\address{Universit\"at Wuppertal, Fachbereich Physik,Gau\ss str.20,
         42097 Wuppertal, Germany
\vspace*{-0.14cm}
}
\author{and the HEGRA collaboration}
\address{complete author list can be found in F. Aharonian {\it et al.},\\
Proc. 24th ICRC, Rome, 1, 474, 1995
\vspace*{-0.14cm}
}
\author {S. Barthelmy, P. Butterworth, T. Cline, N. Gehrels}
\address{NASA/Goddard Space Flight Center, Greenbelt, Maryland 20771
\vspace*{-0.14cm}
}
\author {C. Kouveliotou}
\address{USRA at NASA/Marshall Space Flight Center, Huntsville, Alabama 35812, USA}
\author {J. Fishman, C. Meegan}
\address{NASA/Marshall Space Flight Center, Huntsville, Alabama 35812, USA
}
%%%%%%%%%%%%%%%%%%%%%%%%%%%%%%%%%%%%%%%%%%%%%%%%%%%%%%%%%%%%%%
%
\maketitle\abstracts
{ 
The HEGRA experiment is an air shower detector system for the study 
of neutral and charged cosmic rays in the energy range  
between 500 GeV to 10 PeV.
Here we give an overview of how the
HEGRA detector is used to search for TeV $\gamma$-radiation associated 
with Gamma-Ray Bursts (GRBs) registered with the Burst And Transient Source Experiment (BATSE) 
on board the Compton Gamma-Ray Observatory. 
Furthermore, results of an archival search for GRB radiation above %$E_\gamma>$
15 TeV carried out with the HEGRA air shower arrays  are shown. 
We conclude with a summary of the search activities planned for the future.
}
\small
\vspace*{-.7cm}
\small 
\vspace*{-.2cm}
\section{Introduction} 
\vspace*{-.3cm}
Despite the fact that Gamma-Ray Bursts (GRBs) have been known for more than 
25 years their physical origin, energy and distance scale remain unsolved.~\cite{Fish:95}
Counterparts at any wavelength would help to clarify the situation.
The data taken with the two instruments BATSE \cite{Meeg:92} and
EGRET \cite{Thom:93} on board the Compton Gamma-Ray Observatory 
indicate that TeV counterpart observation is possible:
\begin{itemize}
\item Extrapolations of some of the BATSE \cite{Band:93} and EGRET \cite{Ding:95}
       spectra
       yield large TeV fluxes which could be registered with high significance by  
       present day air shower experiments.
\item  EGRET has observed photons with
            energies up to $18\,$GeV.~\cite{Hurl:94}
\end{itemize}
Also several model scenaria  of burst origin (e.g. the fireball scenario \cite{Mesz:93}
or the scenario of evaporating black holes~\cite{Halz:91}) predict 
TeV emission.
TeV counterpart  observations would impose limits on the GRB distance scale
because the intergalactic infrared radiation field limits the mean free path 
of TeV photons to distances of $\approx$ 100 Mpc.~\cite{Mann:96,Steck:96} 
Moreover TeV detection would reduce the number of possible
emission mechanism models.\\
\newpage
\vspace*{-0.2cm}
In this contribution we describe the searches for TeV counterparts of GRBs
performed with the HEGRA detector system  (La Palma, 
$28.8^\circ$N, $17.9^\circ$W, $2200\,$m above sea level).
The experiment~\cite{Fons:95.1} covers  a system of 6 Cherenkov telescopes,
an array of 245 scintillation counters,
an array of 72 wide angle Cherenkov counters called ''AIROBICC''
and an array of 17 Geiger towers.
The searches can be divided into two classes:
 \begin{enumerate}
        \item \vspace*{-.05cm} {\bf Searches in archival data:} Due to the large field of view
        of the air shower arrays of $>$1 sr, GRBs occur inside the field of view
        of the arrays at a rate of $\approx$25 per year.
        TeV radiation in coincidence with, before and after the GRB detection
        by BATSE can be searched for.
\item \vspace*{-.05cm} {\bf Follow up observations:} The Cherenkov telescopes have a small field of view of
     5 msr. A few minutes after  a  BATSE burst detection the BACODINE~\cite{Bart:95} (BATSE
        Coordinate Distribution Network) distributes preliminary 
        burst coordinates via Internet.
        By slewing the Cherenkov telescopes immediately towards the burst
        direction, follow up observations are started within  only a few minutes after a BATSE
        burst detection.
\end{enumerate}
% After a description of the HEGRA experiment in Section \ref{s0} a
\vspace*{-.05cm} In Section \ref{s1} an overview of the TeV counterpart searches is given. The results of an
archival search performed with  the extensive air shower arrays are described in Section
\ref{s2}.  The activities planned for the future are summarized in Section
\ref{s3}.\\
Recent searches for TeV counterparts with other detectors which did not yield
convincing evidence for TeV GRB  emission can be found in 
\ea{Aglietta}{1992}~\cite{Agli:92}, \ea{Alexandreas}{1994}~\cite{Alex:94}, 
\ea{Allen}{1995}~\cite{Alle:95}, \ea{Amenomori}{1995}~\cite{Amen:95},
\ea{Connaughton}{1995}~\cite{Conn:95} and \ea{Kieda}{1995}~\cite{Kied:96}.
%~\cite {Kied:96,Alex:94,Agli:92,Alle:95,Amen:95,Conn:95}
%
%
%
\vspace{-.35cm}
\section{Overview of the Counterpart Searches Performed  with 
the HEGRA Detector}
\label{s1}
\vspace*{-.2cm}
\subsection*{The Search  in Archival Data with the Air Shower Arrays}
\vspace*{-.1cm}
{\small The scintillator array~\cite{Krawcz:96} registers secondary particles created
 by energetic primaries in the atmosphere.
The array of wide angle Cherenkov  counters 
AIROBICC~\cite{Karl:95.1}  samples the atmospheric Cherenkov light created  by the
secondary particles. 
The scintillator array (number in brackets refer to the AIROBICC array) 
is characterized by a detection threshold for primary photons of 
20 TeV (15 TeV), a mean angular resolution of $\sigma_{63\%}$ =
0.9\deg (0.34\deg) and a duty cycle of 0.95 (0.1) and is
operational since 1989 (1992). 
Gamma/hadron separation information can be inferred by combining the information measured with
scintillator and AIROBICC array~\cite{Prah:95} or by using the data taken with the array of Geiger
towers~\cite{West:95.1}. 
\\
Since the launch of the Compton Gamma Ray Observatory in  January
1991 until June 1996 some  140  BATSE GRBs occurred within the field of view of the
HEGRA arrays.
In a first pioneering study TeV emission from 82 GRBs registered by BATSE  
between January 1991 and April 1994 was searched for in archival
scintillator array data.~\cite{Math:94} 
The search centered on the registration of emission coincident with the
keV/MeV radiation measured by BATSE. 
No positive evidence was found but 90\% confidence level upper limits on the coincident
integral flux above detection threshold energies of $\approx$50 TeV 
between \fluss{-9} and \fluss{-7}  could be established.\\
Recent work has focused on the analysis of BATSE GRBs recorded after April 1994 
with extended and optimized methods.
We have searched for coincident emission as well as for short and long
time emission in the hours preceding and following a burst (see Section \ref{s2}).
The search sensitivity has been increased by optimizing the methods for small number 
statistics and by exploiting  gamma-hadron separation information.~\cite{Krawcz:96.2}
In a related study, short time TeV excesses  independent 
of BATSE GRBs are searched for.~\cite{Padi:95}
In another approach~\cite{Funk:96.1} 
delayed TeV radiation  is searched for which could originate from EeV
burst particles  initiating particle  cascades and thus TeV photons in interactions
with the 2.7\deg K radiation background~\cite{Waxm:96}.
\vspace*{-.25cm}
\subsection*{The Follow Up Searches with the System of Cherenkov Telescopes}
\vspace*{-.2cm}
Cherenkov telescopes measure the atmospheric Cherenkov light created by extensive air
showers.~\cite{Week:88} 
HEGRA has operated two telescopes, the first one~\cite{Mirz:94} since 1992
and the second one~\cite{Kono:96} since 1994.
During this year the system is being extended to its planned full scale.
End of 1996 the installation of the full HEGRA telescope system will be completed and
will consist of one ''prototype telescope'' and 5  ''system telescopes''.~\cite{Pant:95}
A single system telescope is characterized by a detection threshold for 
inducing photons of 500 GeV, a mean angular resolution for
individual events of $\sigma_{63\%}$ = 0.15\deg, a gamma-hadron separation with a 
photon acceptance of 0.55 and hadron rejection of 0.95, a field of view of
5 msr and a duty cycle of 0.1.~\cite{Ulri:95}
\\
The system has been used for follow up observations since February 1995. 
A search is initiated if a burst's
intensity exceeds 1500 counts/sec (February 1995 - March 1996 the threshold was set to 2500 counts/sec), 
which corresponds to a BACODINE position uncertainty of $\sigma\uka$10\deg.
Still, due to the telescopes' small field of view,
the probability  for a single telescope to observe the true GRB position is only 0.05.
Between February 1995 and July 1996 three follow up searches were initiated on
GRBs 950401, 960425 and 960528 (YYMMDD). 
Observations were started approx. 10min after the BATSE detections. The
first two observations were carried out with one, the last observation with 3 telescopes.
For all three GRBs the the Compton/Ulysses  Interplanetary Network of Satellites IPN
~\cite{Hurl:95} has now issued   more precise location estimates which show
that the HEGRA telescopes did not observe the true GRB locations.
\vspace*{-.4cm}
\section{An Archival Search with the Air Shower Arrays}
\vspace*{-.3cm}
\label{s2}
\begin{figure}[t]
\vspace{-.3cm}
  \begin{minipage}[t]{5.7cm}
    \setlength{\epsfxsize}{5.7cm}
    \setlength{\epsfysize}{4.8cm}
           \mbox {\epsffile{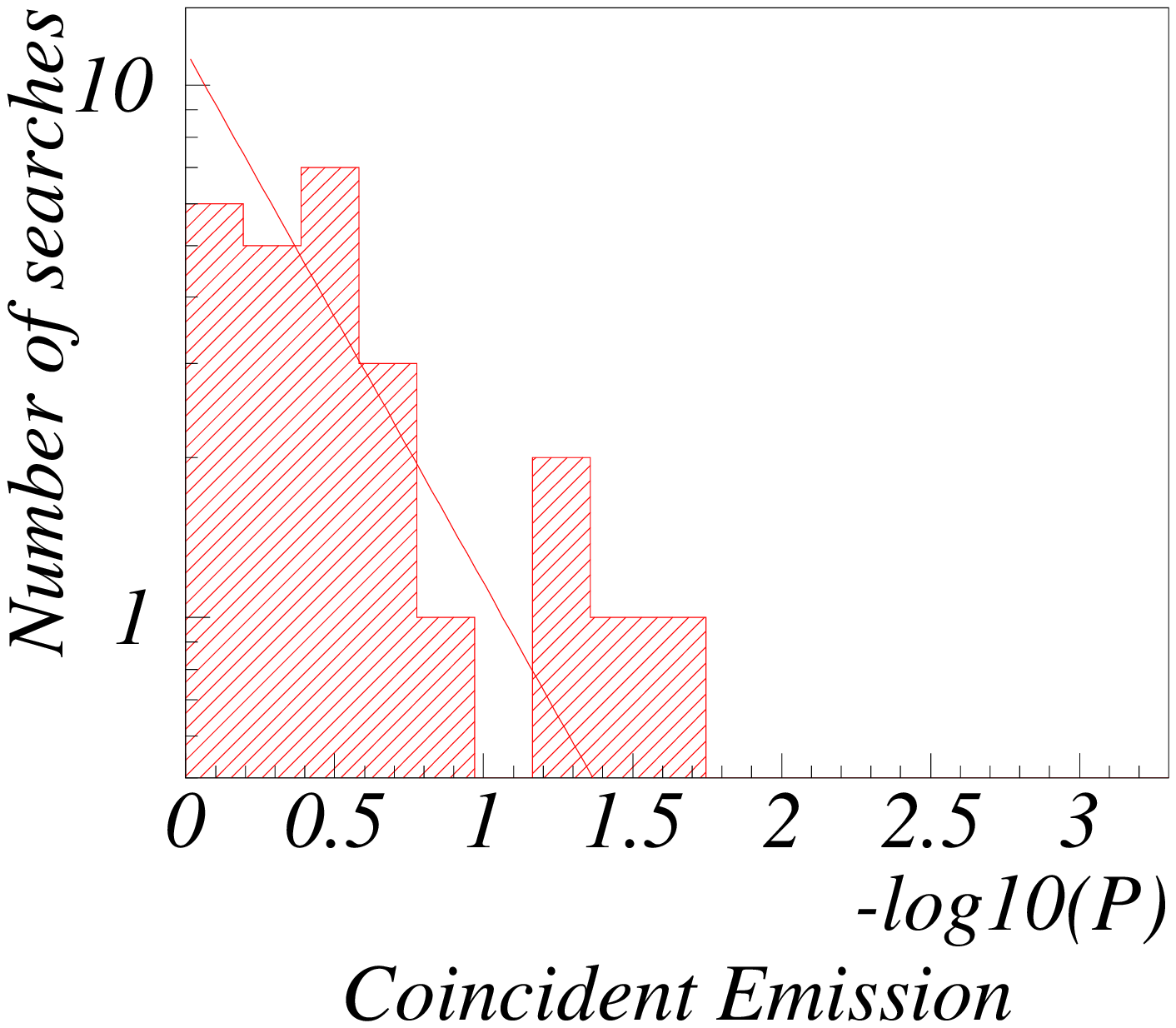}}
 \end{minipage} \hfill
 \begin{minipage}[t]{5.7cm}
    \setlength{\epsfxsize}{5.7cm}
    \setlength{\epsfysize}{4.8cm}
           \mbox {\epsffile{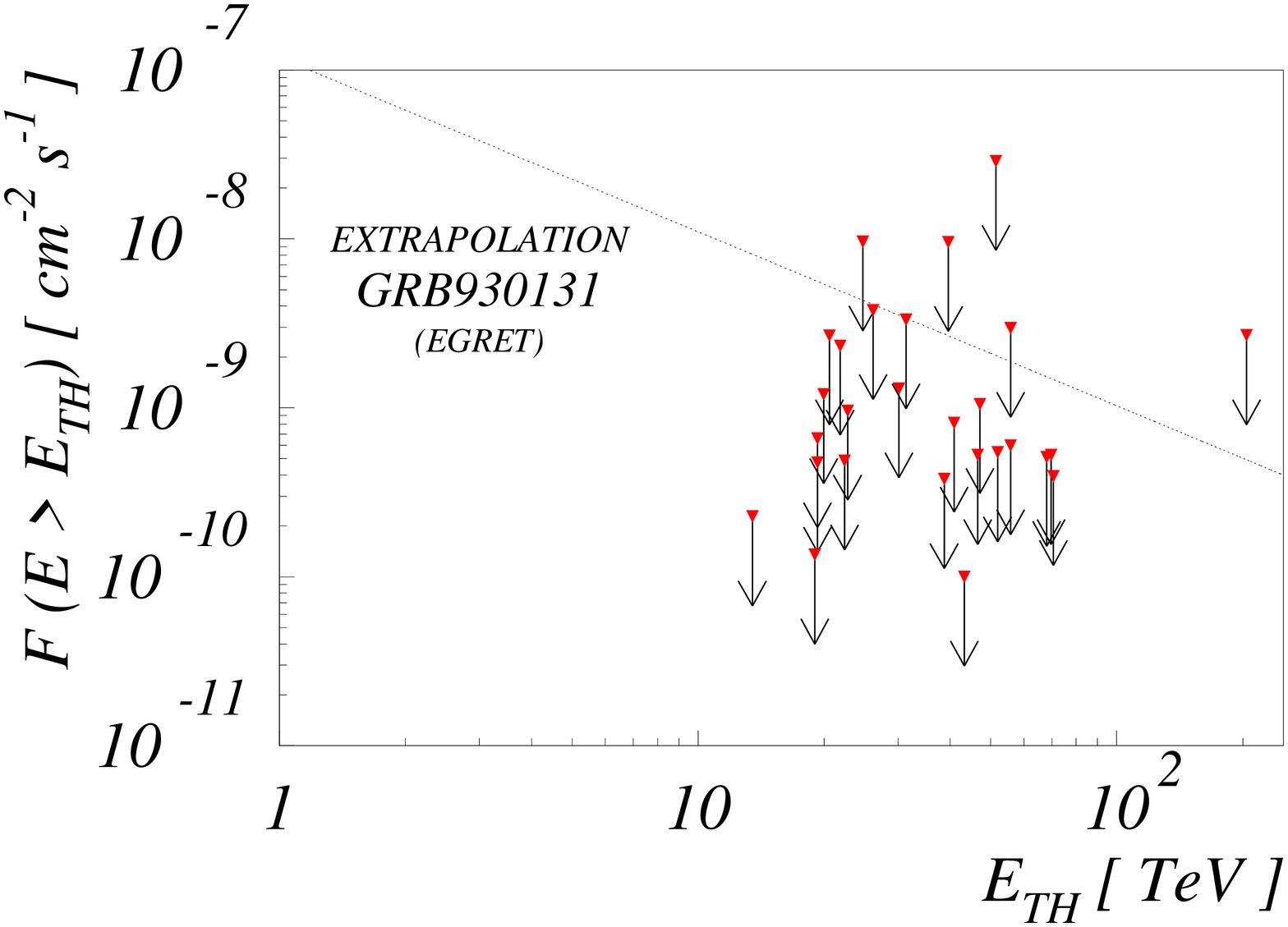}}
  \end{minipage}
\vspace*{-.6cm}
  \caption{ \label {coinc}   
                              Results of the searches for coincident emission of 27 bursts.
                              On the left side the distribution of the chance probabilities $p$ is shown.
                              The line depicts the distribution expected for the case that
                              there are no counterparts in the data.
                              On the right side the upper limits on the coincident flux 
                              (90\% confidence level)
                              versus the threshold energy for inducing photons are shown.
                              The threshold energy is a function of the zenith 
                              angle of the location of a burst.
                              To guide the eye an extrapolated GRB 
                              spectrum \protect\cite{Somm:94} is displayed.
}
\vspace*{-.55cm}
\end{figure}
In this section the preliminary results of a search for TeV radiation in archival
scintillator array and AIROBICC array data are described.
Thirty four GRBs recorded with BATSE
between 1993 and 1995 were studied.
The locations and durations of eighteen bursts were taken from the 3B catalog and 
the data of thirteen bursts have been transmitted to us by the BACODINE facility.
For each burst three searches were performed:
\begin{itemize}
   \item A search for {\bf coincident emission}, i.e. for radiation within
         the time interval BATSE registered 90\% of the burst counts.
   \item A search for {\bf short time emission} with $60\,$sec intervals, two successive
         intervals overlapping half a minute.
   \item A search for {\bf long time emission} with time intervals of
         a length of $2^i\,$min for integer $i\ge0$.  
\end{itemize}
\vspace*{-.05cm}
The intervals of the second and third search covered the
time in which the burst location had been in the 
field of view of the HEGRA experiment ($\theta<30^\circ$).
For each time interval either the $2\,\sigma$ error region given 
by BATSE or the smaller ring segment given by the Compton/Ulysses Interplanetary Network of
Satellites were searched for emission.
In order to exploit the better angular resolution of the HEGRA arrays in comparison
to BATSE, grids of optimal search bins were used. Repeating the search with shifted grids
ensured that a large fraction of about $90\%$ of hypothetical source photons would 
fall inside  one of the bins.
With the trigger rate and the detector acceptance in horizontal coordinates
the background was calculated ~\cite{Alex:92} and Monte Carlo data were generated.
A burst candidate was the most significant excess found in a search.
With the help of the Monte Carlo data
the {\bf chance probability ${p}$} for the background to yield a
burst candidate equally or more significant than the observed one was calculated.\\
The search for coincident emission was applied to 
27, the search for short time emission to 26 and 
the search for long time emission to 23 out of the 34 bursts,
depending on the position of a GRB in the local sky 
and on interruptions of the detector exposure due to calibration runs.
The following results have been obtained:
\begin{figure}[t]
\vspace*{-0.3cm}
  \begin{minipage}[t]{5.7cm}
    \setlength{\epsfxsize}{5.7cm}
    \setlength{\epsfysize}{4.8cm}
           \mbox {\epsffile{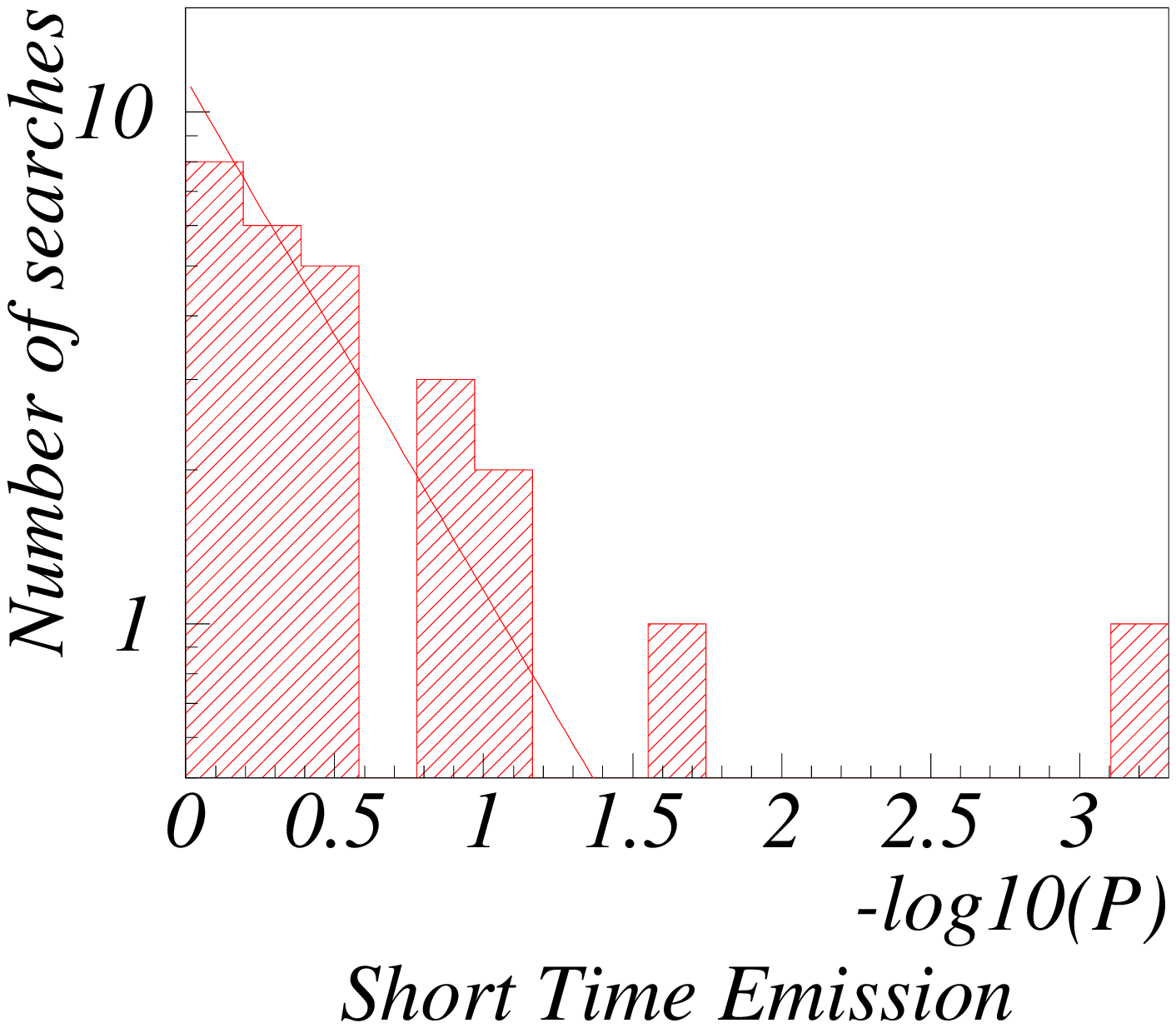}}
 \end{minipage} \hfill
 \begin{minipage}[t]{5.7cm}
    \setlength{\epsfxsize}{5.7cm}
    \setlength{\epsfysize}{4.8cm}
           \mbox {\epsffile{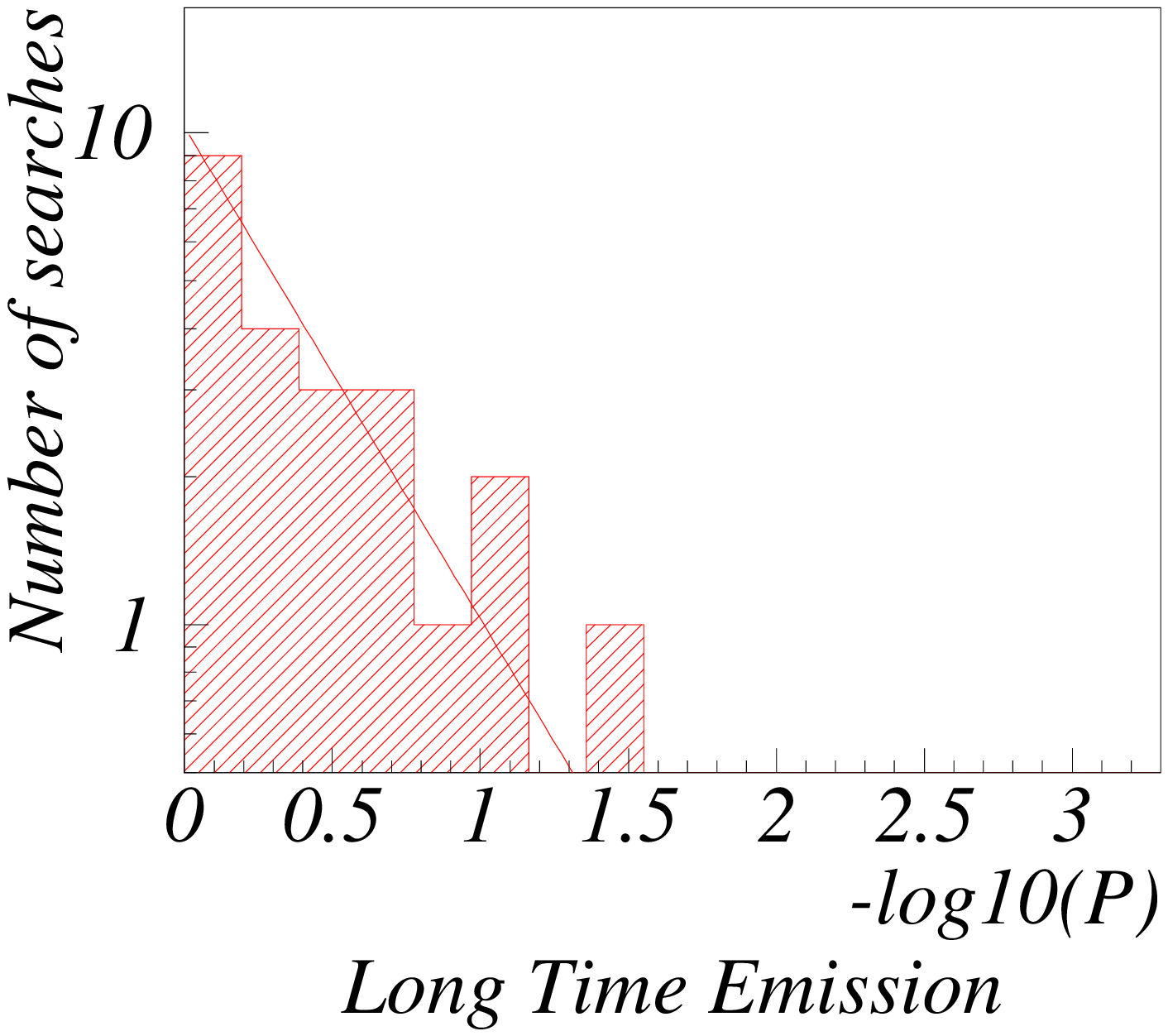}}
  \end{minipage}
\vspace*{-0.6cm}
    \caption{ \label{cp.eps}  Distributions of the chance probabilities $p$ for the
                              searches for short and long time emission.
                              The lines show the distributions expected if
                              there are no counterparts.}
\vspace*{-0.4cm}
\end{figure}
\begin{itemize}
 \item {\bf Coincident emission:} 
       In Figure \ref{coinc} (left side) the chance probabilities $p$ of the searches are shown.
        The observations are consistent with the background expectation.
       In Figure \ref{coinc} (right side) the upper limits on the observed flux (90\% confidence level) 
       are shown.
       The upper limits are well within the region of extrapolated GRB spectra. 
 \item {\bf Short time emission:}
        A remarkable excess (Fig. \ref{cp.eps}, left side) was found for GRB 950701 (06:35 UTC). 
        During one minute, 86 minutes before the burst triggered the BATSE detector,
        12 events were observed from the burst direction where 1.7 were expected.
        The significance of the raw excess is 6$\sigma$. Taking into account all search bins
        in temporal and solid angle space the significance of this excess is $3.4\sigma$.
 \item {\bf Long time emission:}
        No significant excess has been found
        (Fig. \ref{cp.eps}, right side).
\end{itemize}
\vspace*{-.05cm}%
Altogether the 76 searches yield results which are consistent with a non-detection
of TeV counterparts.
\vspace*{-.2cm}
\section{Future Activities}
\vspace*{-.2cm}
\label{s3}
The described searches with the air shower arrays and the system of Cherenkov
Telescopes will be continued.\\
The follow up searches will be extended:
\begin{enumerate}
\item Follow up observations will be performed not only with the Cherenkov
Telescopes but also with the scintillator array.  After a burst warning an ''angle sensitive
trigger'' will lower the threshold condition of 14 counters in coincidence to 3
counters in coincidence, while only accepting showers from a 20\deg FWHM centered
on the burst direction. By this means the energy threshold of the scintillator
array is lowered from 20 TeV to 3 TeV while increasing the trigger rate only slightly.
\item Follow up observations will be initiated not only by BATSE via BACODINE but
also by the instrument HETE which will be launched on October 1996. For this
purpose a HETE receiver will be installed on the HEGRA site. 
\end{enumerate}
\vspace*{-.3cm}
\section*{Acknowledgments} 
\vspace*{-.1cm}
This work was supported by the Bundesministerium f\"ur Bildung und Forschung
BMBF  under contract numbers 05 2HH 264 and 05 2WT 164, the Deutsche
Forschungsgemeinschaft DFG and the Spanish Research Council CYCIT. We
thank the Instituto de Astroph\'isica de Canarias IAC for supplying
excellent working conditions at La Palma.
\vspace*{-.3cm}
\section*{References}
\vspace*{-.1cm}
\end{document}